\documentclass{elsart}
\usepackage{epsfig}
\usepackage{enumerate}

\begin{document}
\begin{frontmatter}

\title{Phase transition in kinetic exchange opinion models with independence}

\author{Nuno Crokidakis}
\thanks{nuno@if.uff.br}

\address{
Instituto de F\'{\i}sica, \hspace{1mm} Universidade Federal Fluminense \\
Av. Litor\^anea s/n, \hspace{1mm} 24210-340 \hspace{1mm} Niter\'oi - RJ, \hspace{1mm} Brazil}

\maketitle

\begin{abstract}
\noindent
In this work we study the critical behavior of a three-state ($+1$, $-1$, $0$) opinion model with independence. Each agent has a probability $q$ to act as independent, i.e., he/she can choose his/her opinion independently of the opinions of the other agents. On the other hand, with the complementary probability $1-q$ the agent interacts with a randomly chosen individual through a kinetic exchange. Our analytical and numerical results show that the independence mechanism acts as a noise that induces an order-disorder transition at critical points $q_{c}$ that depend on the individuals' flexibility. For a special value of this flexibility the system undergoes a transition to an absorbing state with all opinions $0$.

\end{abstract}
\end{frontmatter}


\section{Introduction}

In the recent years, the statistical physics techniques have been successfully applied in the description of socioeconomic phenomena. Among the studied problems we can cite opinion dynamics, language evolution, biological aging, dynamics of stock markets, earthquakes and many others \cite{galam_book,sen_book,pmco_book}. These interdisciplinary topics are usually treated by means of computer simulations of agent-based models, which allow us to understand the emergence of collective phenomena in those systems.

Recently, the impact of nonconformity in opinion dynamics has atracted attention of physicists \cite{galam,lalama,sznajd_indep1,sznajd_indep2,sznajd_indep3}. Anticonformists are similar to conformists, since both take cognizance of the group norm. Thus, conformers agree with the norm, anticonformers disagree. On the other hand, we have the independent behavior, where the individual tends to resist to the groups' influence. As discussed in \cite{sznajd_indep2,sznajd_indep3}, independence is a kind of nonconformity, and it acts on an opinion model as a kind of stochastic driving that can lead the model to undergo a phase transition. In fact, independence plays the role of a random noise similar to social temperature \cite{lalama,sznajd_indep2,sznajd_indep3}. 

In this work we study the impact of independence on agents' behavior in a kinetic exchange opinion model. For this purpose, we introduce a probability $q$ of agents to make independent decisions. Our analytical results and numerical simulations show that the model undergoes a phase transition at critical points $q_{c}$ that depend on another model parameter, related to the agents' flexibility.

This work is organized as follows. In Section 2 we present the microscopic rules that define the model and in Section 3 the numerical and analytical results are discussed. Finally, our conclusions are presented in Section 4.


\section{Model}

Our model is based on kinetic exchange opinion models (KEOM) \cite{lccc,p_sen,biswas11,biswas}. A population of $N$ agents is defined on a fully-connected graph, i.e., each agent can interact with all others, which characterizes a mean-field-like scheme. In addition, each agent $i$ carries one of three possible opinions (or states), namely $o_{i}=+1$, $-1$ or $0$. The following microscopic rules govern the dynamics:

\begin{enumerate}

\item An agent $i$ is randomly chosen;

\item With probability $q$, this agent will act independently. In this case, with probability $g$ he/she chooses the opinion $o_{i}=0$, with probability $(1-g)/2$ he/she adopts the opinion $o_{i}=+1$ and with probability $(1-g)/2$ he/she chooses the opinion $o_{i}=-1$;

\item On the other hand, with probability $1-q$ we choose another agent, say $j$, at random, in a way that $j$ will influence $i$. Thus, the opinion of the agent $i$ in the next time step $t+1$ will be updated according to
\begin{equation}\label{eq1}
o_{i}(t+1) = {\rm sgn}\left[o_{i}(t) + o_{j}(t)  \right]\,,
\end{equation}
where the sign function is defined such that ${\rm sgn}(0)=0$.
\end{enumerate}

In the case where the agent $i$ does not act independently, the change of his/her state occur according to a rule similar to the one proposed recently in a KEOM \cite{biswas}. Notice, however, that in Ref. \cite{biswas} two randomly chosen agents $i$ and $j$ interact with competitive couplings, i.e., the kinetic equation of interaction is $o_{i}(t+1) = {\rm sgn}\left[o_{i}(t) + \mu_{ij}\,o_{j}(t)  \right]$. In this case, the couplings $\mu_{ij}$ are random variables presenting the value $-1$ ($+1$) with probability $p$ ($1-p$). In other words, the parameter $p$ denotes the fraction of negative interactions. In this case, the model of Ref. \cite{biswas} undergoes a nonequilibrium phase transition at $p_{c}=1/4$. In the absence of negative interactions ($p=0$), the population reaches consensus states with all opinions $+1$ or $-1$. 

Thus, our Eq. (\ref{eq1}) represents the KEOM of Ref. \cite{biswas} with no negative interactions, and the above parameter $g$ can be related to the agents' flexibility \cite{sznajd_indep1}. In this case, for $q=0$ (no independence) all stationary states will give us $O=1$, where $O$ is the order parameter of the system, 
\begin{equation} \label{eq2}
O = \left\langle \frac{1}{N}\left|\sum_{i=1}^{N}\,o_{i}\right|\right\rangle ~, 
\end{equation}
and $\langle\, ...\, \rangle$ denotes a disorder or configurational average taken at steady states. The Eq. (\ref{eq2}) defines the ``magnetization per spin'' of the system. We will show by means of analytical and numerical results that the independent behavior works as a noise that induces a phase transition in the KEOM in the absence of negative interactions. 

The three states considered in the model can be interpreted as follows \cite{vazquez1,vazquez2,lalama2}. We have a population of voters that can choose among two candidates A and B. Thus, the opinions represent the intention of an agent to vote for the candidate A (opinion $+1$), for the candidate B (opinion $-1$), or the agent may be undecided (opinion $0$). In this case, notice that there is a difference among the undecided and independent agents. An agent $i$ that decide to behave independently (with probability $q$) can make a decision to change or not his/her opinion based on his/her own conviction, whatever is the his/her current state $o_{i}$ (decided or undecided). In other words, an interaction with an agent $j$ is not required. On the other hand, an undecided agent $i$ can change his/her opinion $o_{i}$ in two ways: due to an interaction with a decided agent $j$ (following the rule given by Eq. (\ref{eq1}), with probability $1-q$) or due to his/her own decision to do that (independently, with probability $q$).

Regarding the independent behavior, one can consider the homogeneous case ($g=1/3$) and the heterogeneous one ($g\neq 1/3$). These cases will be considered separately in the next section.


\section{Results}

\subsection{Homogeneous case: $g=1/3$}

One can start studying the homogeneous case $g=1/3$. In this case, we have that all probabilities related to the independent behavior, namely $g$ and $(1-g)/2$, are equal to $1/3$. Thus, the probability that an agent $i$ chooses a given opinion $+1$, $-1$ or $0$ independently of the opinions of the other agents is $q/3$. For the analysis of the model, we have considered the order parameter $O$ defined by Eq. (\ref{eq2}), as well as the susceptibility $\chi$ and the Binder cumulant $U$ \cite{binder,binder2}, defined as
\begin{eqnarray} \label{eq3}
\chi & = &  N\,(\langle O^{2}\rangle - \langle O \rangle^{2}) \, ~ \\ \label{eq4}
U &  = &  1 - \frac{\langle O^{4}\rangle}{3\,\langle O^{2}\rangle^{2}} \,.
\end{eqnarray}

\begin{figure}[t]
\begin{center}
\vspace{3mm}
\includegraphics[width=0.45\textwidth,angle=0]{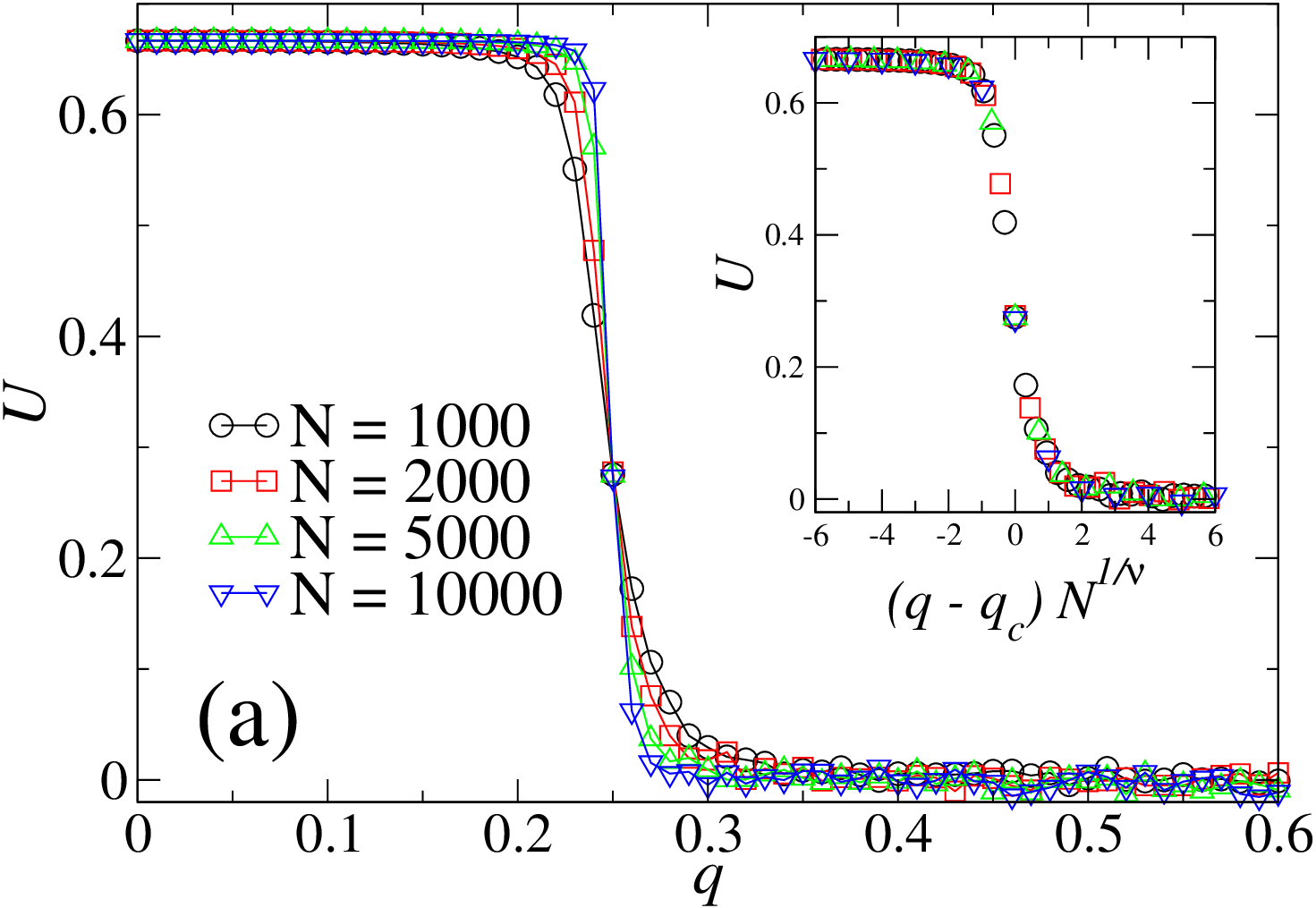}
\hspace{0.4cm}
\includegraphics[width=0.45\textwidth,angle=0]{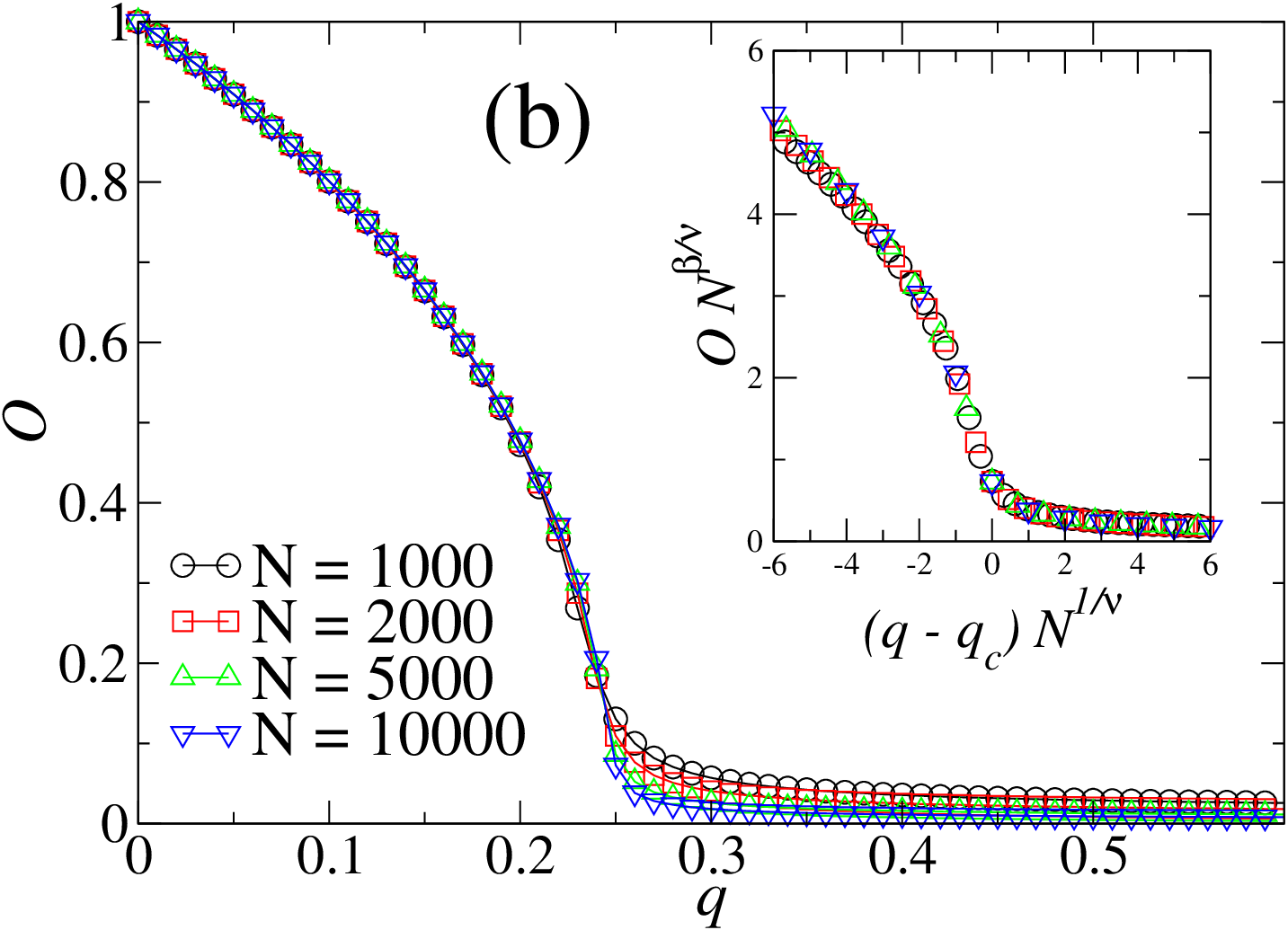}
\\
\vspace{0.8cm}
\includegraphics[width=0.45\textwidth,angle=0]{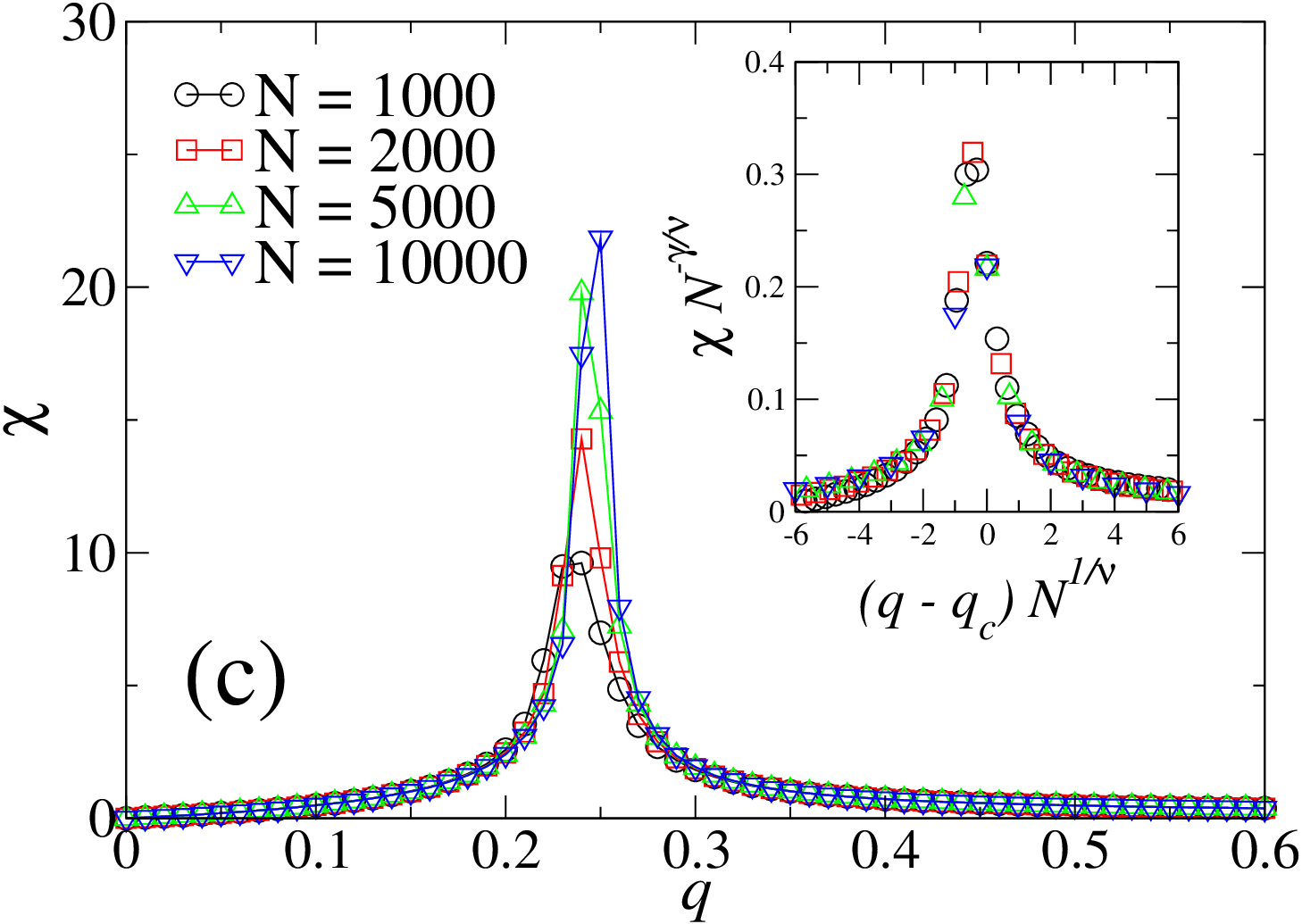}
\end{center}
\caption{(Color online) Binder cumulant $U$ (a), order parameter $O$ (b) and susceptibility $\chi$ (c) as functions of the independence probability $q$ for the homogeneous case ($g=1/3$) and different population sizes $N$. In the inset we exhibit the corresponding scaling plots. The estimated critical quantities are $q_{c}\approx 0.25$, $\beta\approx 0.5$, $\gamma\approx 1.0$ and $1/\nu\approx 0.5$. Results are averaged over $300$, $250$, $200$ and $150$ samples for $N=1000, 2000, 5000$ and $10000$, respectively.}
\label{fig1}
\end{figure}

Notice that the Binder cumulant defined by Eq. (\ref{eq4}) is directly related to the order's parameter \textit{kurtosis} $k$, that can be defined as $k=\langle O^{4}\rangle/3\,\langle O^{2}\rangle^{2}$. The initial configuration of the population is fully disordered, i.e., we started all simulations with an equal fraction of each opinion ($1/3$ for each one). In addition, one time step in the simulations is defined by the application of the rules defined in the previous section $N$ times. In Fig. \ref{fig1} we exhibit the quantities of interest as functions of $q$ for different population sizes $N$. All results suggest the typical behavior of a phase transition. In order to estimate the transition point, we look for the crossing of the Binder cumulant curves for the different sizes \cite{binder,binder2,salinas}. From Fig. \ref{fig1} (a), the estimated value is $q_{c}=0.25 \pm 0.002$, which agrees with the analytical prediction $q_{c}=1/4$ [see Eq. (\ref{qc_sym}) of the Appendix]. In addition, in order to determine the critical exponents associated with the phase transition we performed a finite-size scaling (FSS) analysis. We have considered the standard scaling relations, 
\begin{eqnarray} \label{eq5}
O(N) & \sim & N^{-\beta/\nu} \\  \label{eq6}
\chi(N) & \sim & N^{\gamma/\nu} \\   \label{eq7}
U(N) & \sim & {\rm constant} \\   \label{eq8}
q_{c}(N) - q_{c} & \sim & N^{-1/\nu} ~,
\end{eqnarray}
that are valid in the vicinity of the transition. Thus, we exhibit in the insets of Fig. \ref{fig1} the scaling plots of the quantities of interest ($U$, $O$ and $\chi$). Our estimates for the critical exponents are $\beta\approx 0.5$, $\gamma\approx 1.0$ and $1/\nu\approx 0.5$. Notice that the critical probability, $q_{c}=1/4$, presents the same value of the critical fraction of negative interactions ($p_{c}=1/4$) of the KEOM of Ref. \cite{biswas}. In addition, the critical exponents are the same in the two formulations of the model. Thus, the inclusion of the independent behavior with equal probabilities (i.e., $g=1/3$) produces a similar effect to the introduction of negative interactions in the KEOM of Ref. \cite{biswas}.


\subsection{Heterogeneous case: $g\neq 1/3$}

\begin{figure}[t]
\begin{center}
\vspace{0.5cm}
\includegraphics[width=0.45\textwidth,angle=0]{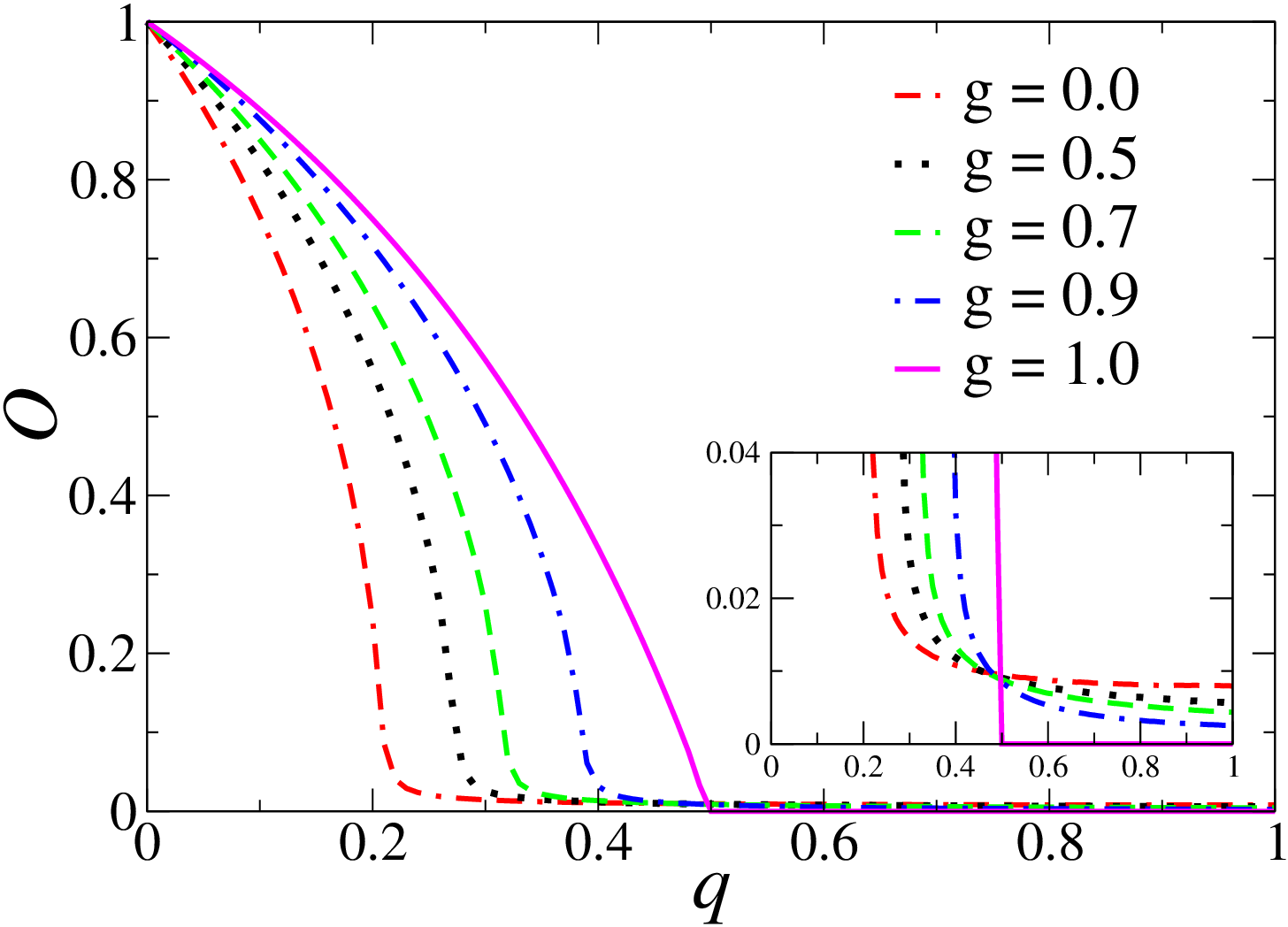}
\end{center}
\caption{(Color online) Order parameter $O$ as a function of $q$ for $N=10000$ and typical values of $g$. One can see that the transition points depend on $g$. The inset shows the region near $O=0$. Results are averaged over $150$ simulations.}
\label{fig2}
\end{figure}

One can also consider the general case where $g\neq 1/3$. In this case, for an agent that act independently, the probabilities to choose the three possible opinions are different. As in the previous subsection, we started all simulations with an equal fraction of each opinion. In Fig. \ref{fig2} we show the order parameter as a function of $q$ for typical values of $g$ and population size $N=10000$. One can see that the phase transition occurs for all values of $g$ exhibited in Fig. \ref{fig2}, and the critical points depend on $g$, i.e., we have $q_{c}=q_{c}(g)$. Furthermore, another interesting result that one can see in Fig. \ref{fig2} is that for $g=1$ the order parameter goes exactly to $O=0$ at $q_{c}\approx 0.5$, presenting no finite-size effects as the other curves do, as can be easily seen in the inset of Fig. \ref{fig2}. This fact can be easily understood. Indeed, for $g=1$ all agents that behave independently choose the opinion $o=0$. Thus, for a sufficiently large value of $q$ all agents will change independently to $o=0$. In this case, Eq. (\ref{eq2}) give us an order parameter $O=0$. This qualitative discussion can be confirmed by analytical considerations (see the Appendix).

\begin{figure}[t]
\begin{center}
\vspace{1.0cm}
\includegraphics[width=0.45\textwidth,angle=0]{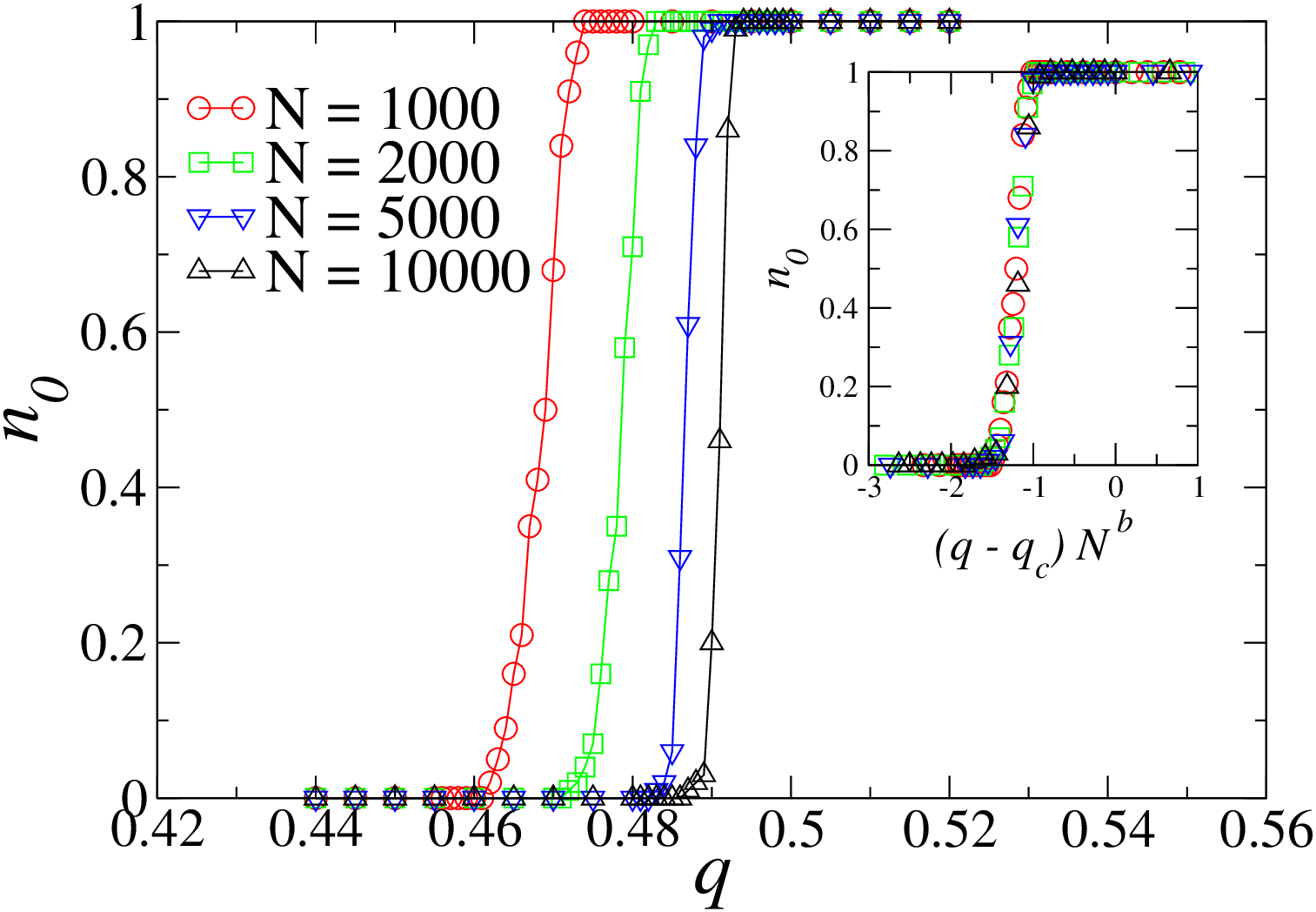}
\end{center}
\caption{(Color online) Fraction $n_{0}$ of samples (over $200$ simulations) that reaches the $0$ consensus as a function of $q$ for $g=1$ and typical population sizes $N$ (main plot). In the inset it is exhibited the corresponding scaling plot. The best collapse of data was obtained for $q_{c}=0.5$ and $b=0.53$.}
\label{fig3}
\end{figure}

Thus, the case $g=1$ is special, because all agents change their opinions to $o=0$ for a sufficient large value of the parameter $q$. Indeed, if all agents are in the $o=0$ state, the evolution equation (\ref{eq1}), when applied (with probability $1-q$), does not change the opinions to $+1$ or $-1$ anymore, which means that the system is in an absorbing state. This fact, together with the absence of finite-size effects for the order parameter defined in Eq. (\ref{eq2}), suggests that one can not apply the scaling relations (\ref{eq5}) - (\ref{eq8}) for $g=1$. In this case, it is better to analyze other quantity as an order parameter, as was done, for example, for the 2D Sznajd model \cite{stauffer,adriano}. Thus, following \cite{stauffer,adriano}, we performed several simulations of the system for $g=1$ and we measured the fraction $n_{0}$ of samples that reached the absorbing state with all opinions $o=0$ as a function of $q$. The result is exhibited in Fig. \ref{fig3} for typical values of $N$, and in this case this order parameter strongly depends on the system size. Considering scaling relations in a similar way as in Ref. \cite{adriano}, i.e., plotting $n_{0}$ as a function of the variable $(q - q_{c})\,N^{b}$, one obtains $q_{c}=0.50 \pm 0.003$, in agreement with the previous discussion, and $b=0.53 \pm 0.02$. The corresponding data collapse is exhibited in the inset of Fig. \ref{fig3}.

\begin{figure}[t]
\begin{center}
\vspace{1.0cm}
\includegraphics[width=0.5\textwidth,angle=0]{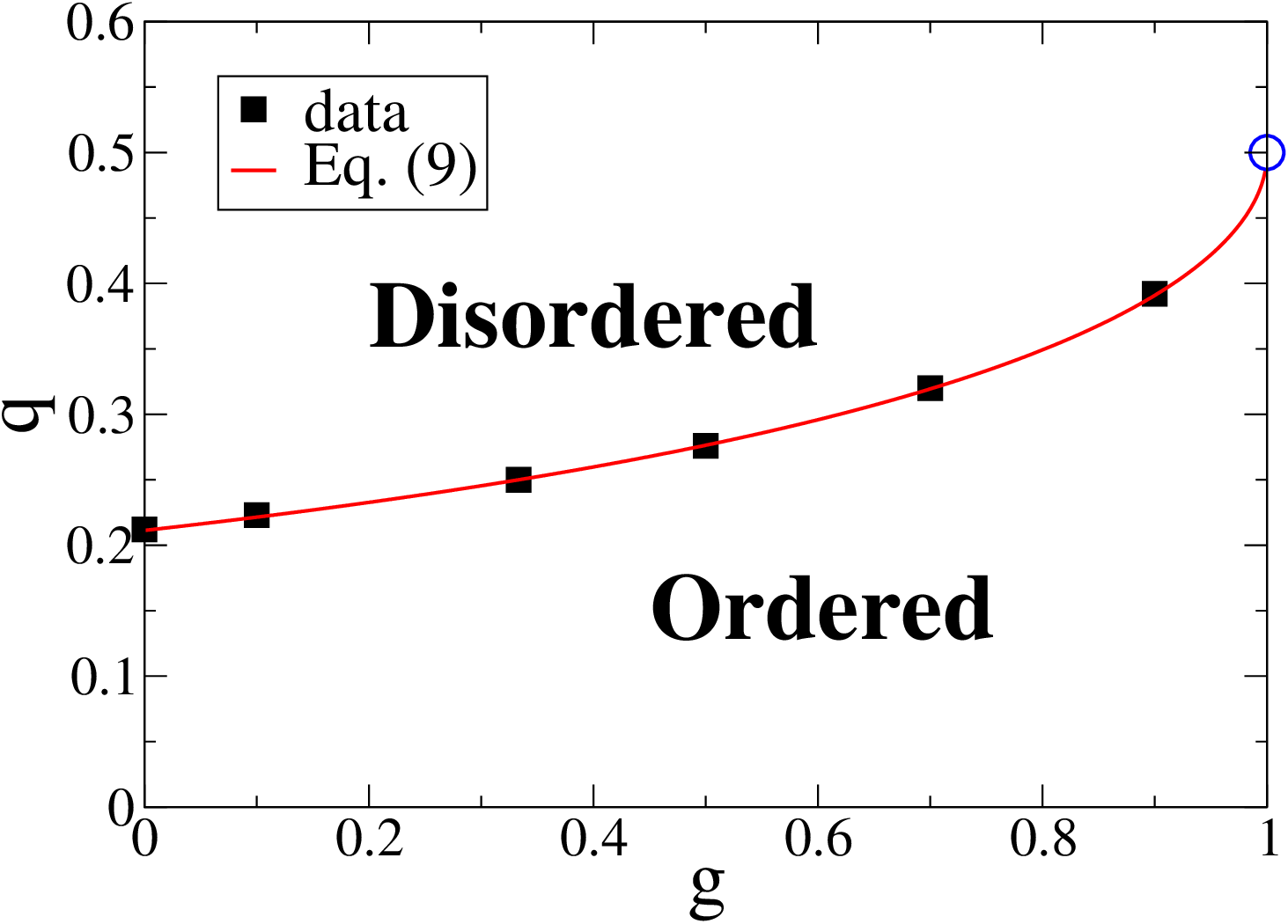}
\end{center}
\caption{(Color online) Phase diagram of the model in the plane $q$ versus $g$, separating the ordered and the disordered phases. The symbols are the numerical estimates of the critical points $q_{c}$, whereas the full line is the analytical prediction, Eq. (\ref{eq9}). The open (blue) circle denotes the special case $g=1$, as discussed in the text. The error bars determined by the FSS analysis are smaller than data points.}
\label{fig4}
\end{figure}

As above discussed, the numerical results suggest that critical points $q_{c}$ depend on $g$. This picture is confirmed by the analytical solution of the model, which give us (see Eq. (\ref{qcs}) of the Appendix)
\begin{equation} \label{eq9}
q_{c}(g) = \frac{1}{2}\,\left[1 - \left(\frac{1-g}{3-g}\right)^{1/2} \right] \, .
\end{equation}
\noindent
Notice that the above solution give us $q_{c}(g=1/3) = 1/4$, and the exact result for $g=1$ is $q_{c}(g=1)=1/2$, which agrees with the above discussion. We performed a FSS analysis based on Eqs. (\ref{eq5}) - (\ref{eq8}) in order to obtain the critical points and the critical exponents for other values of $g<1$. In Fig. \ref{fig4} the Eq. (\ref{eq9}) is plotted together with all numerical estimates of $q_{c}(g)$. One can see that the numerical results agree very well with the analytical prediction. In addition, the critical exponents are the same for all values of $g<1$, i.e., we have $\beta\approx 0.5$, $\gamma\approx 1.0$ and $1/\nu\approx 0.5$, which indicates a universality on the order-disorder frontier of the model, except on the ``special'' point $g=1$.

\section{Final remarks}   

In this work we introduce the mechanism of independence in a three-state ($+1$, $-1$ and $0$) kinetic exchange opinion model. In the absence of negative interactions, this model always evolve to ordered (consensus) states. Our results show that independence acts as a noise, inducing a nonequilibrium phase transition in the model, and that the critical points depend on the agents' flexibility $g$. The numerical simulations suggest that we have the same critical exponents for all values of $g<1$, i.e., we have $\beta\approx 0.5$, $\gamma\approx 1.0$ and $1/\nu\approx 0.5$, which indicates a universality on the order-disorder frontier of the model. This is an expected result, due to the mean-field character of the interactions. On the other hand, the case $g=1$ is special, and the system undergoes a phase transition to an absorbing state with all agents in the undecided state $o=0$.


\appendix
\section{Appendix} 
\label{app}

Following the lines of Refs. \cite{biswas11,biswas}, we computed the critical values of the probability $q$. We first obtained the matrix of transition probabilities whose elements $m_{i,j}$ furnish the probability that a state suffers the shift or change $i \to j$. Let us also define $f_{1}$, $f_{0}$ and $f_{-1}$, the stationary probabilities of each possible state. In the steady state, the fluxes into and out from a given state must balance. In particular, for the  null state, one has
\begin{equation} \label{nullstate}
m_{1, 0}+m_{-1, 0}=m_{0,1}+m_{0,-1}   \,.
\end{equation}

Moreover, when the order parameter vanishes, it must be $f_1=f_{-1}$. Finally, let us define $r(k)$, with $-2\le k \le 2$, the probability that the state shift per unit time is $k$, that is, $r(k)=\sum_i m_{i,i+k}$. In the steady state, the average shift must vanish, namely, 
\begin{equation} \label{nullshift}
 2[r(2)-r(-2)] +r(1)-r(-1)=0 \,.
\end{equation}

For the more general case considering the flexibility parameter $g$, the elements $m_{i,j}$ of the transition matrix are
 
\begin{eqnarray} \nonumber
m_{1, 1} &=& (1 - q)\,f_1^{2} + \frac{(1-g)}{2}\,q\,f_{1} \\ \nonumber
m_{1, 0} &=& (1-q)\,f_{1}\,f_{-1} + g\,q\,f_{1}     \\ \nonumber
m_{1, -1} &=& \frac{(1-g)}{2}\,q\,f_{1} \\ \nonumber
m_{0, 1} &=& (1-q)\,f_{0}\,f_{1} + \frac{(1-g)}{2}\,q\,f_{0}   \\ \nonumber
m_{0, 0} &=& (1-q)\,f_{0}^{2} + g\,q\,f_{0}  \\ \nonumber
m_{0, -1} &=& (1-q)\,f_{0}\,f_{-1} + \frac{(1-g)}{2}\,q\,f_{0}  \\ \nonumber
m_{-1, 1} &=& \frac{(1-g)}{2}\,q\,f_{-1} \\ \nonumber
m_{-1, 0} &=&  (1-q)\,f_{1}\,f_{-1} + g\,q\,f_{-1} \\ \nonumber
m_{-1, -1} &=& (1-q)\,f_{-1}^{2} + \frac{(1-g)}{2}\,q\,f_{-1}  
\end{eqnarray}

First, one can consider the homogeneous case $g=1/3$. In this case, the above elements $m_{i,j}$ are simplified, and the null state condition (\ref{nullstate}) give us $f_1=f_{-1}=f_0=1/3$ (disorder condition). Thus, the null average shift condition (\ref{nullshift}), together with the above disorder condition, leads to 
 
\begin{equation} \label{qc_sym}
q_{c} = \frac{1}{4} \,.
\end{equation}
 
For the more general case, the conditions (\ref{nullstate}) and (\ref{nullshift}) lead to a second-order equation for the variable $q_{c}$,
\begin{equation} \label{qc_non_sym}
6\,(3-g)\,q_{c}^{2} + 2\,(3g-9)\,q_{c} + 3 = 0  \,,
\end{equation}
which give us two distinct solutions, namely 
\begin{eqnarray} \label{qcs}
q_{c}^{\pm} =  \frac{1}{2}\,\left[1 \pm \left(\frac{1-g}{3-g}\right)^{1/2} \right] \, .
\end{eqnarray}

Although both solutions are mathematically valid, the solution $q_{c}^{+}$ leads to $f_{0}>1$ in the disordered phase, and consequently $f_{1}<0$ and $f_{-1}<0$. On the other hand, the solution $q_{c}^{-}$ is physically acceptable because it leads to $f_{0}<1$ as well as $f_{0}<1$ and $f_{0}<1$, satisfying the normalization condition $f_{1} + f_{-1} + f_{0} = 1$. Thus, the physically valid analytical solution for the general model is given by $q_{c}^{-}$. In particular, we have $q_{c}=1/4$ for $g=1/3$ [which agrees with Eq. (\ref{qc_sym})] and $q_{c}=1/2$ for $g=1$. In addition, it can be shown that the null state condition (\ref{nullstate}) for $g=1$ give us the solution $f_{0}=1$ in the disordered phase, and then $f_{1}=f_{-1}=0$. This explains the result $O=0$ for $q\geq 1/2$ observed in Figs. \ref{fig2} and \ref{fig3}, that was discussed in Section 3.


\section*{Acknowledgments}

The author acknowledges financial support from the Universidade Federal Fluminense, Brazil.

\end{document}